\documentclass[twocolumn,showpacs,preprintnumbers,amsmath,amssymb,aps,prd,letterpaper,floatfix,nofootinbib,superscriptaddress,10pt]{revtex4-2}

\usepackage[T1]{fontenc}
\usepackage[utf8]{inputenc}
\usepackage[UKenglish]{babel}
\usepackage{mathtools,hyperref,tikz,physics,subcaption,dsfont}
\usepackage{booktabs}
\usepackage{array}
\usepackage[justification=raggedright]{caption}

\begin{document}
\begin{flushright}
    \texttt{DESY-25-115}
\end{flushright}

\title{Current-Enhanced Excited States in Lattice QCD Three-Point Functions}
\author{Lorenzo Barca}
\email{lorenzo.barca@desy.de}
\affiliation{John von Neumann-Institut für Computing (NIC), Deutsches Elektronen-Synchrotron DESY, Platanenallee 6, 15738 Zeuthen, Germany}
\begin{abstract}
Excited-state contamination remains one of the leading sources of systematic uncertainty in the precise determination of hadron structure observables from lattice QCD.
In this letter, we present a general argument, inspired by current–meson dominance and implemented through the variational method, 
to identify which excited states are enhanced by the choice of the inserted current and kinematics.
The argument is supported by numerical evidence across multiple hadronic channels and 
provides both a conceptual understanding and practical guidance to account for excited-state effects in hadron three-point function analyses.
\end{abstract}

\maketitle

{\textbf{Introduction.---}}
Lattice Quantum Chromodynamics (Lattice QCD) provides a first-principles framework to compute a wide range of hadronic observables by numerically simulating the strong interactions of quarks and gluons. 
In particular, hadron three-point correlation functions offer valuable insight into the internal structure of hadrons and their interactions mediated by external currents. 
These functions encode how a hadron responds to the insertion of a current—probing, for instance, its form factors or charge distributions—while accounting for the full complexity of QCD dynamics.
The standard expression for a hadronic three-point function is
\begin{equation}
\label{c3pt}
C_{\rm 3pt}(\vec{p}', t; \vec{q}, \tau) = 
\langle \mathrm{O}_H(\vec{p}', t) ~\mathcal{J}(\vec{q}, \tau) ~ \mathrm{O}^\dagger_H(\vec{p}, 0) \rangle~,
\end{equation}
where $\mathrm{O}_H$ and $\bar{\mathrm{O}}_H$ are interpolators that annihilate and create hadrons with the quantum numbers of $H$, and $\mathcal{J}$ is a current operator that mediates the interaction. 
For simplicity, we consider the case where the same hadron operator appears at both source and sink, and the current is a local bilinear operator. 
However, the general argument applies more broadly. The spectral decomposition of the three-point function exposes its dependence on the matrix elements of the current between hadronic states. Explicitly, one finds
\begin{align}
\nonumber
\label{spectral_c3pt}
C_{\rm 3pt}(\vec{p}', t; \vec{q}, \tau) &= Z'_H Z_H^\dagger ~
\langle H'|\mathcal{J}| H \rangle e^{-E'_H(t-\tau)} e^{-E_H\tau}
\\
+
\sum_{f}
\sum_{i}
&Z'_f Z_i^\dagger ~\langle f' |\mathcal{J}| i \rangle
e^{-E'_f(t-\tau)} e^{-E_i\tau}~,
\end{align}
where the prefactors $Z$ arise from the overlaps of the interpolating operators with the physical states. The ground-state contribution is shown explicitly, while the double sum accounts for all possible remaining excited initial and/or final states with the same quantum numbers as $H$.
In the asymptotic regime $t\gg \tau \gg 0$, contributions from excited states are exponentially suppressed if  $E_H' < E_f'$ or $E_H < E_i$. In this limit, the ground state matrix element can be, in principle, isolated. However, in practice, the signal-to-noise ratio of baryonic correlators deteriorates exponentially with increasing time separation, complicating the extraction of hadron properties.
Consequently, lattice QCD analyses are constrained to intermediate source–sink separations where excited-state contamination remains significant, making its understanding essential.

Notably, long before QCD, current algebra and meson dominance models provided powerful predictions for hadron dynamics.
Vector meson dominance (VMD), proposed by Sakurai \cite{Sakurai:1960ju}, postulates that the electromagnetic current couples to hadrons predominantly via the $\rho$, $\omega$, and $\phi$ mesons.
Based on VMD, quantitative predictions were made for processes such as $\rho\to e^+e^-$ decay \cite{Sakurai1969}, $\omega \to \pi^0\gamma$ and $\phi\to \eta\gamma$ \cite{Kroll:1967it,Oakes:1967zz}, as well as $e^+e^-\to\pi^+\pi^-,\pi^+\pi^-\pi^0$ and the $q^2$-dependence of pion form factors \cite{Gell-Mann:1961jim}; see \cite{Meissner:1987ge} for a review.
Similarly, current algebra and the Weinberg sum rules, together with vector and axial-vector meson dominance \cite{Weinberg:1967kj}, yield the empirical relation $m_{a_1}/m_\rho \approx \sqrt{2}$, in striking agreement with experiment.
The partially conserved axial current (PCAC) relation further explains the strong coupling of axial currents to pions, reflecting their role as pseudo-Goldstone bosons of spontaneously broken chiral symmetry, and underpins soft-pion theorems, pion pole dominance in form factors, and low-energy current-induced pion production \cite{Adler:1964um,Adler:1965ga,Altarelli:1971jh,Bernard:1995dp}.

Drawing on the intuition from current–meson dominance, one can predict which multi-hadron states are preferentially produced in a given channel, provided the relevant symmetries and kinematics permit it. These \emph{current-enhanced} states can dominate excited-state contamination in lattice QCD correlators, even if their masses exceed those of other states. In this work, we use the variational method to show that these contributions originate from a \emph{volume enhancement} of specific excited states generated by quark-line disconnected diagrams, and that only certain operators can produce them. Combining this with current–meson dominance arguments yields clear, physically intuitive predictions for which states dominate in a given three-point function. In the nucleon channel and in $B\to\pi$ decays, these predictions agree with Chiral Perturbation Theory (ChPT) expectations, and they extend naturally to cases without ChPT guidance. Our framework is general, applicable to any hadron three-point (or higher), and we present numerical evidence confirming these predictions.

{\textbf{Volume enhancement of current-enhanced states.---}}To show the current-enhanced excited-state contributions, we employ the variational method with an extended basis of single and multi-hadron operators that have the quantum numbers of the initial and final operators:
$    \mathbb{B}_{H} = 
    \{ \mathrm{O}_{H}, 
    \mathrm{O}_{H_{1}}, 
    \mathrm{O}_{H_{2}},
    \mathrm{O}_{H M},
    \mathrm{O}_{H_{1} M_1}, ... \}
$.
Using this basis, we construct a matrix of two-point correlation functions, which reads
\begin{equation}
C_{\rm 2pt}(\vec{p}, t)_{kj}
=
\langle \mathrm{O}_{k}(\vec{p}, t)~\mathrm{O}^\dagger_{j}(\vec{p}, 0) \rangle
\qquad
\mathrm{O}_{k, j} \in \mathbb{B}_{H}~.
\end{equation}
We solve the Generalised Eigenvalue Problem (GEVP)
\begin{equation}
\label{gevp_c2pt}
C_{\rm 2pt}(\vec{p}, t) V(\vec{p}, t_0) = C_{\rm 2pt}(\vec{p}, t_0) \Lambda(\vec{p}, t) V(\vec{p}, t)
\qquad t>t_0~,
\end{equation}
which yields the matrix of eigenvalues $\Lambda(\vec{p}, t)=\mathrm{diag}\left(\lambda^H,\lambda^{H^*}, \lambda^{HM}, ...\right)$ and eigenvectors $V(\vec{p}, t)=\left(\vec{v}_{\mathrm{O}_H}, \vec{v}_{\mathrm{O}_{H_1}}, \vec{v}_{\mathrm{O}_{H_2}}, \vec{v}_{\mathrm{O}_{HM}}, \vec{v}_{\mathrm{O}_{H_1M_1}},...\right)$. 
The eigenvectors can be used to diagonalise the system \cite{Blossier:2009kd} 
and to construct an improved operator $\widetilde{\mathrm{O}}_{H}$ like
\begin{equation}
\widetilde{\mathrm{O}}_{H} = \sum_{\mathrm{O}_i \in \mathbb{B}_H} v^{H}_{\mathrm{O}_i} \mathrm{O}_i~.
\end{equation}
By construction, this operator overlaps only with the state $H$, assuming the variational basis spans the full Hilbert space.
However, in practice, it is not feasible to include the infinite tower of interpolating operators required to fully reconstruct the spectrum.
Due to this limitation, one must work with a truncated basis, and with the right choice of $t_0$ and $t$ in Eq.~\eqref{gevp_c2pt}, 
the associated systematic effects are exponentially suppressed by the energy gaps between the omitted states and the resolved ones \cite{Bulava:2011yz}.
Using this diagonalised operator, we can compute the improved three-point functions
\begin{equation}
\label{c3pt_imp}
\widetilde{C}_{\rm 3pt}(\vec{p}', t; \vec{q}, \tau) = 
\langle \widetilde{\mathrm{O}}_H(\vec{p}', t) ~\mathcal{J}(\vec{q}, \tau) ~ \widetilde{\mathrm{O}}^\dagger_H(\vec{p}, 0) \rangle~.
\end{equation}
This expression involves a linear combination of three-point correlators built from all operators in the variational basis. Specifically, it requires evaluating correlation functions of the form
\begin{equation}
\langle \mathrm{O}_{k}(\vec{p}', t)~\mathcal{J}(\vec{q}, \tau) ~\mathrm{O}^\dagger_j(\vec{p}, 0) \rangle
\qquad
\mathrm{O}_{k,j} \in \mathbb{B}_{H}.
\end{equation}
Among these are three-point correlators of the type
\begin{equation}
\label{c3pt_ohm}
\langle 
\mathrm{O}_{H}(\vec{p}'_{H} t)\mathrm{O}_{M}(\vec{p}'_{M} t)
~\mathcal{J}(\vec{q}, \tau)
~\mathrm{O}^\dagger_{H}(\vec{p}, 0) \rangle~,
\end{equation}
with $\vec{p}'=\vec{p}'_H + \vec{p}'_M$. 
These terms, as we argue below, can be significantly enhanced for specific choices of current, 
momenta, and interpolating operators.
To illustrate the enhancement mechanism, we rewrite the three-point function in Eq.~\eqref{c3pt_ohm} in terms of the Wick contractions obtained from the quark content of the hadron operators: 
\begin{align}
\label{c3pt_multiparticle}
\langle 
\mathrm{O}_{H}(\vec{p}'_{H} t)\mathrm{O}_{M}(\vec{p}'_{M} t)
~\mathcal{J}(\vec{q}, \tau)
~ \mathrm{O}^\dagger_{H}(\vec{p}, 0) \rangle
&=
\\
\nonumber
\langle W_C(\vec{p}'_{H}; \vec{p}'_{M}; \vec{q})
+
W_D(\vec{p}'_{H}; \vec{p}'_{M}; \vec{q})
+
W_{\rm disc}&(\vec{p}'_{H}; \vec{p}'_{M}; \vec{q}) \rangle
~.
\end{align}
In lattice QCD, $W_C$ denotes the \textit{connected} diagrams, $W_D$ the \textit{direct} diagrams, and $W_{\rm disc}$ the \textit{disconnected} diagrams.
Both $W_D$ and $W_{\rm disc}$ are \textit{quark-line disconnected}.
Which contributions appear, and with what relative weight, depends on the quark flavor structure of the interpolating operators, the current, and the kinematics.
To make these contributions explicit, we write each term in coordinate space, including the appropriate Fourier transforms to project onto the desired momenta:
\onecolumngrid
\begin{align}
\label{connected}
W_C(\vec{p}'_{H}; \vec{p}'_{M}; \vec{q}) 
&=
\sum_{\vec{x}_{H}, \vec{x}_{M}, \vec{z}}
e^{-i\vec{p}'_{H} \cdot \vec{x}_{H}}
e^{-i\vec{p}'_{M} \cdot \vec{x}_{M}}
e^{i\vec{q}\cdot \vec{z}}
~
\langle
\mathrm{O}_{H}(\vec{x}_{H}, t)\mathrm{O}_{M}(\vec{x}_{M}, t)
~\mathcal{J}(\vec{z}, \tau)
~ \mathrm{O}^\dagger_{H}(\vec{x}_0, 0) 
\rangle_F
\\
\label{direct}
 W_D(\vec{p}'_{H}; \vec{p}'_{M}; \vec{q}) 
&=
\sum_{\vec{x}_{H}, \vec{x}_{M}, \vec{z}}
e^{-i\vec{p}'_{H} \cdot \vec{x}_{H}}
e^{-i\vec{p}'_{M} \cdot \vec{x}_{M}}
e^{i\vec{q}\cdot \vec{z}}
~
\langle \mathrm{O}_{H}(\vec{x}_{H}, t) ~ \mathrm{O}^\dagger_{H}(\vec{x}_0, 0)  \rangle_F
~
\langle \mathrm{O}_{M}(\vec{x}_{M}, t)~\mathcal{J}(\vec{z}, \tau)\rangle_F
\\
\label{disconnected}
W_{\rm disc}(\vec{p}'_{H}; \vec{p}'_{M}; \vec{q}) 
&=
\sum_{\vec{x}_{H}, \vec{x}_{M}, \vec{z}}
e^{-i\vec{p}'_{H} \cdot \vec{x}_{H}}
e^{-i\vec{p}'_{M} \cdot \vec{x}_{M}}
e^{i\vec{q}\cdot \vec{z}}
~
\langle \mathrm{O}_{H}(\vec{x}_{H}, t) ~ \mathrm{O}^\dagger_{H}(\vec{x}_0, 0)  \rangle_F
~\langle \mathrm{O}_{M}(\vec{x}_{M}, t)\rangle_F
~\langle \mathcal{J}(\vec{z}, \tau)\rangle_F~.
\end{align}
\twocolumngrid
Here, $\langle \cdot \rangle_F$  denotes the Wick contractions, expressed as traces over products of quark propagators.
Examples of the connected and direct diagrams are shown in Fig.~\ref{fig:wick} in the Appendix, 
for the case of a baryon-meson two-hadron operator at the sink and a baryon interpolator at the source.

A central observation of this work is that,
when the meson momentum matches the current momentum transfer, $\vec{p}'_M=\vec{q}$,
the amplitude of the quark-line disconnected contributions--such as those in Eqs.~\eqref{direct}-\eqref{disconnected}--
is significantly larger than the quark-line connected contribution in Eq.~\eqref{connected}.
In the non-interacting limit, such contributions are expected to be enhanced by the spatial volume ($L^3/a^3$), since translational invariance can be used for the current-meson two-point functions.
In several nucleon channels—axial-vector, vector, and scalar—we find that at $\tau = t$ the quark-line disconnected contributions are $\mathcal{O}(100)$ larger than the connected ones on an ensemble with $L=24a=2.4~\mathrm{fm}$ and $m_\pi = 429~\rm MeV$.
By comparing results from two volumes \cite{Barca:2024hrl}, $L=24a=2.4~\mathrm{fm}$ and $L=48a=4.2~\mathrm{fm}$, we confirm that this enhancement scales with the spatial volume, in agreement with theoretical expectations in the non-interacting limit.

Therefore, in the improved three-point functions of Eq.~\eqref{c3pt_imp}
and at the moderate intermediate distances currently accessible ($\tau, ~t-\tau \lesssim 0.8~\rm fm$ for nucleon observables), 
operators whose correlation functions include quark-line disconnected contributions
dominate over those that do not.
This simple but important observation has direct implications:
the presence (or absence) of quark-line disconnected diagrams in a given three-point function
indicates which excited states are most likely to contribute significantly at accessible distances.
This connection becomes especially transparent in a partially quenched framework \cite{PhysRevD.102.074502}, where individual contractions can be mapped to specific intermediate states in the spectral decomposition.

For instance, the meson–current term appearing in Eq.~\eqref{direct} contains the matrix element $\langle 0|\mathcal{J}|M\rangle$, describing the creation of a meson by the current.
The quantum numbers of the meson $M$ can be inferred from current algebra,
providing theoretical guidance on which states couple most strongly to a given current.
Importantly, although the quark-line disconnected diagrams are large in amplitude,
their time dependence reflects the propagation of multi-particle states,
and leads to a faster exponential decay compared to ground-state contributions present in connected diagrams.

This behavior becomes clear in the schematic structure of the connected and direct terms:
\begin{align}
\nonumber
W_C & =
Z'_{HM} \langle (HM)'|\mathcal{J} | H \rangle Z^\dagger_{H} e^{-E'_{HM}(t-\tau)} e^{-E_{H}\tau} 
\\
&+|Z'_{H}| \langle H'|\mathcal{J} | H \rangle Z^\dagger_{H} e^{-E'_{H}(t-\tau)} e^{-E_{H}\tau} + 
\dots
\\
W_D & = \mathcal{O}(L^3/a^3) ~|Z_H|^2 |Z_{M}|^2 e^{-E_{H}t} e^{-E_{M}(t-\tau)} + \dots
\end{align}
The connected contributions $W_C$ contain both single- and multi-particle intermediate states,
as propagation from source to sink can proceed through either type.
In contrast, $W_D$ is dominated by multi-particle propagation and thus decays more rapidly in Euclidean time.
Consequently, although quark-line disconnected terms can be $\mathcal{O}(100)$ larger in amplitude, they are exponentially suppressed at large time separations relative to the ground-state contribution in $W_C$.
For the source-sink separations currently accessible in lattice QCD, however,
these enhanced contributions dominate the signal and must be properly accounted for.
This behavior is supported by numerical evidence reported in \cite{Barca:2024sub, Alexandrou:2024tin},
and is consistent with the qualitative picture from partially quenched theory \cite{PhysRevD.102.074502}, which associates multi-hadron contributions with specific diagram topologies.

Importantly, the states created by the current in these diagrams are not limited to stable single hadrons.
In lattice QCD, resonances such as the $\rho$ or $\sigma$ mesons do not appear as asymptotic states due to their strong decays.
Instead, the current often couples directly to their decay products: for instance, the vector current can produce $\pi\pi$ in P-wave, while the scalar current can create $\pi\pi$ in S-wave.
It is important to keep in mind that the dominant states can depend sensitively on the ensemble parameters—such as the pion mass and spatial volume—since these determine which multi-hadron states lie closest to threshold.
We provide further discussion and numerical examples of such cases below.

{\textbf{Strategies to account for current-enhanced states.---}}%
\label{sec:strategies}%
The identification of quark-line disconnected diagrams as the dominant contributions at short and intermediate source-sink separations provides a valuable insight:
once these diagrams are known to drive the contamination, one can develop targeted strategies to account for the corresponding excited states.

This can be tackled through these proposed strategies:
 
\begin{itemize}
\item {\it Variational analysis with tailored operators}:
Include interpolators with strong overlap onto the enhanced multi-particle states, such as $N\sigma$ \cite{Barca:2024hrl}, $N\pi$ \cite{Barca:2022uhi, Alexandrou:2024tin}, or $N\rho$-like operators (see App.~\ref{app:vector_varanalysis}).
This model-independent approach isolates and subtracts specific excited-state contributions, and enables determination of transition matrix elements, e.g. $\langle N\pi| \mathcal{J} | N\rangle$ \cite{Barca:2024sub}.
Recent results show that including only operators producing current-enhanced states already yields accurate and robust outcomes.

\item {\it Two-point analysis of meson–current diagrams}:
Compute the quark-line disconnected diagrams—or just the current–meson two-point function—to extract the energy of the associated multi-particle state and identify which states are current-enhanced.
These states can be then incorporated into effective field theory to compute analytic expressions for their contributions, or used as priors in multi-state fits, or to directly remove contamination in standard three-point functions \cite{Aoki:2025taf, Tsuji:2025quu}.

\item {\it Effective field theory estimates}:
Use low-energy EFTs, such as ChPT, treating the current-enhanced meson explicitly to compute analytic expressions for their contributions. Information from the current-meson two-point analysis guides which couplings to include in the chiral Lagrangian.
\end{itemize}

While the variational method--particularly with an extended operator basis--offers the most reliable control over current-enhanced states, it is computationally expensive.
The other two strategies are significantly cheaper and—when combined—can still yield robust analytic estimates of excited-state contamination.

Our framework also clarifies why in \cite{Alexandrou:2024tin} the inclusion of $N\pi$ operators in some nucleon channels—such as electromagnetic nucleon form factors or nucleon sigma terms—yields little improvement, since these operators lack quark-line disconnected contributions and are therefore not current-enhanced.

{\textbf{Examples: Numerical evidence in nucleon three-point functions.---}}%
The expression for the nucleon three-point function reads
\begin{equation}
\label{c3pt_nucleon}
C_{\rm 3pt}(\vec{p}', t; \vec{q}, \tau) = 
\langle \mathrm{O}_N(\vec{p}', t) ~\mathcal{J}(\vec{q}, \tau) ~ \mathrm{O}^\dagger_N(\vec{p}, 0) \rangle~,
\end{equation}
whose spectral decomposition contains the nucleon-current matrix elements $\langle N(\vec{p}')|\mathcal{J}(\vec{q})|N(\vec{p})\rangle$.
In the following, we discuss three examples for which we have numerical evidence: axial, scalar and vector channels.

{\it {Nucleon axial matrix elements.---}}%
In the case of an isovector intermediate axial current $\mathcal{J} = \mathcal{A}_\mu = \bar{\psi} \gamma_\mu \gamma_5 \psi$, there is a non-vanishing direct diagram when the meson operator is a pion-like or $a_1$-like interpolator.
In the first case, the direct diagram reads
\begin{equation}
\label{direct_diagram_npi}
W_D
=
\langle \mathrm{O}_{N}(\vec{p}'_{N}, t) \mathrm{O}^\dagger_{N}(\vec{p}, 0)  \rangle_F
 \langle \mathrm{O}_{\pi}(\vec{p}'_{\pi}, t) \mathcal{A}_\mu(\vec{q}, \tau)\rangle_F
\end{equation}
and the gauge average of the pion-axial term yields
\begin{equation}
\label{axial_pion_me}
\langle \langle \mathrm{O}_{\pi}(\vec{p}_{\pi}, t)~\mathcal{A}_\mu(\vec{q}, \tau) \rangle_F \rangle
=
i q_\mu f_\pi \delta_{\vec{p}_\pi, \vec{q}} ~e^{-E_\pi(t-\tau)}~.
\end{equation}
Therefore, this diagram is non-zero only when the pion carries the same momentum of the current and along the direction of its component.
Several lattice studies have reported a large excited-state contamination in this channel for the standard nucleon three-point function in Eq.~\eqref{c3pt_nucleon}, e.g., \cite{Bali:2018qus, Jang:2019vkm, RQCD:2019jai}.
It was then predicted at leading order in ChPT \cite{Bar:2018xyi, Bar:2019gfx, RQCD:2019jai} that $N\pi$ states contribute very largely to such channels,
and that their contribution is not suppressed by the volume, in agreement with the argument proposed in this text.

Furthermore, ChPT predicts that $N\pi$ states contribute only weakly—compared to other channels—to cases with $\vec{q}=\vec{p}'=\vec{p}=\vec{0}$ \cite{Tiburzi:2015tta, Hansen:2016qoz}, which include, for example, the axial charge \cite{Hall:2025ytt}.
This is again in agreement with the fact that the direct diagram in Eq.~\eqref{direct_diagram_npi} vanishes in these kinematics. 
However, in the axial channel $\mathcal{A}_4$ and forward limit with boosted frames, i.e. $\vec{p}'=\vec{p}$, Eq.~\eqref{direct_diagram_npi} does not vanish 
and indeed, both ChPT \cite{RQCD:2019jai} and lattice results \cite{Barca:2022uhi, Alexandrou:2024tin} show a large contamination from $N\pi$ states.

As for the excited-state contamination in the forward limit with $\vec{q}=\vec{p}'=\vec{0}$, we speculate that the axial meson $a_1$ plays a prominent role, 
and in particular we suggest to include $N(\vec{0}) a_1(\vec{0})$ in the variational analysis or account for the excited states as discussed in the previous section.
Very likely, the $N a_1$-like operator will couple well to the scattering states, 
i.e., $N \oplus$ \textit{decays products} of $a_1$ \cite{10.1093/ptep/ptaa104}.

{\it {Nucleon scalar matrix elements.---}}%
The nucleon scalar matrix elements can be extracted from nucleon three-point functions with a scalar current insertion $\mathcal{J}=\mathcal{S}$.
Of particular interest are nucleon isoscalar scalar matrix elements in the forward limit, 
which are proportional to the nucleon scalar coupling and sigma terms \cite{Alexandrou:2024ozj, Agadjanov:2023efe}.

Following the same argument of the previous sections, when $N\sigma$ operators are included in the variational basis, the isoscalar scalar channel can receive sizable contributions from both the direct and the disconnected terms in Eqs.~\eqref{direct}-\eqref{disconnected}.
In this case, the direct and disconnected terms read
\begin{align}
\nonumber
W_D(\vec{p}'_{N}; \vec{p}'_{\sigma}; \vec{0}) 
&=
\langle \mathrm{O}_{N}(\vec{p}'_{N}, t) \mathrm{O}^\dagger_{N}(\vec{0}, 0)  \rangle_F
\\
&\times
\langle \mathrm{O}_{\sigma}(\vec{p}'_{\sigma}, t) \mathcal{S}(\vec{0}, 0)  \rangle_F
\\
\nonumber
W_{\rm disc}(\vec{p}'_{N}; \vec{p}'_{\sigma}; \vec{0}) 
&=
\langle \mathrm{O}_{N}(\vec{p}'_{N}, t) \mathrm{O}^\dagger_{N}(\vec{0}, 0)  \rangle_F
\\
&\times
\langle \mathrm{O}_{\sigma}(\vec{p}'_{\sigma}, t)\rangle_F
\langle \mathcal{S}(\vec{0}, 0) \rangle_F
\end{align}
The specific intermediate states contributing to these terms depend on the ensemble parameters, such as the pion mass and spatial volume.

In the regime where the pion mass is sufficiently large that the low-lying scalar mesons lie below the multi-particle threshold, the physical interpretation is that $N\sigma$ states contribute to the disconnected piece, while $N a_0$ states to the direct piece.
In \cite{Barca:2024hrl}, we verify numerically the large $N\sigma$ contribution in the isoscalar channel at $m_\pi=429$ MeV, where the sigma lies at $m_\sigma =554(49)~\rm MeV$, below the $\pi\pi$ threshold.

At lighter pion masses, where the $\sigma$ is unstable, in the sense that its mass is above the $\pi\pi$ S-wave threshold, the dominant contribution is instead expected from $N\pi\pi$ S-wave states or from mixed $N\sigma$–$N\pi\pi$ components, due to their coupling \cite{Barca:2025prep}.
A next-to-next-to-leading order ($\mathrm{N}^2\mathrm{LO}$) ChPT calculation at the physical point \cite{Gupta:2021ahb} predicts large $N\pi$ and $N\pi\pi$ contributions.
However, in \cite{Alexandrou:2024tin} a variational analysis including the lowest $N\pi$ states finds no effect on this observable—consistent with the absence of quark-line disconnected contributions from $N\pi$ operators in this channel.
We therefore advocate using $N\sigma$-like operators even at light pion masses, since they will couple to the relevant $N\pi\pi$ scattering states.


{\it {Nucleon vector matrix elements.---}}%
The vector matrix elements can be computed by inserting a vector current $\mathcal{J}=\mathcal{V}_\mu$ in Eq.~\eqref{c3pt_nucleon}.
Following the same steps of the previous section, in the isovector channel, we expect a large contamination 
from $N\rho$ states, in the physical scenario where the $\rho$ energy is below the $\pi\pi$ energy threshold.
This may be counterintuitive as the $\rho$ is quite heavy and may be therefore missed in the multi-state fit analyses.
To prove this, we have carried out a variational analysis with $N$ and $N\rho$ operators on an ensemble with $N_f=3$ and $m_\pi=429~\rm MeV$, where the $\rho$ lies at $m_\rho\approx 860~\rm MeV$,
well below the $\pi\pi$.
In App.\ref{app:vector_varanalysis}, we report numerical evidence that shows that by removing $N\rho$ states from the nucleon isovector vector three-point functions, 
the excited-state contamination is exponentially reduced, proving that $N\rho$ states are the dominant contribution.
Notice that on this ensemble, the energy of the $N\rho$ states with zero and unit total momemtum is 
$E_{N\rho}\approx 2280~\rm MeV$, and $2170~\rm MeV$, respectively -- substantially higher than the corresponding energies of $N\sigma$, $N\pi$, or $N\pi\pi$ states.

This analysis delivers an important message: current-enhanced multi-particle states can dominate the excited-state contamination even if they are heavy. 
While such states might appear to be suppressed due to their large energy, their strong overlap with the current via the quark-line disconnected diagrams,
allow them to contribute significantly at moderate Euclidean time separations. 
In particular, the $N\rho$ contribution—even when heavier—can outcompete lighter states such as $N\pi$, due to its stronger coupling to the vector current.
In \cite{Alexandrou:2024ozj}, a variational analysis with $N$- and $N\pi$-like operators shows that the $N\pi$ states have little effect in this channel, again in perfect agreement with our argument.

Importantly, at sufficiently light pion masses, where the $\rho$ becomes unstable, 
we expect that the dominant contribution is coming from $N\pi\pi$ with $\pi\pi$ in P-wave in the rest frame, 
or a mixture of $N\rho$ and $N\pi\pi$.

{\textbf{Other examples: Semileptonic decays.---}}%
The argument presented in this work is general and extends naturally to processes where different hadrons appear at the source and sink, such as in semileptonic decays.

In the case of heavy-light decays, like, for instance {\it {$B\to \pi \ell {\nu}_\ell$}}, 
the expression for the three-point function reads
\begin{equation}
\langle \mathrm{O}_{B}(\vec{0}, t) ~\mathcal{V}_\mu(\vec{q}, \tau)~\mathrm{O}^\dagger_\pi(-\vec{q}, 0)\rangle~.
\end{equation}
For such correlation functions, the direct diagrams arise from two-hadron operators of the type $B_\mu^* \pi$,
which can have the quantum numbers of $B$.
In particular, the expression is
\begin{align}
\nonumber
W_D(\vec{0}; \vec{p}'_{\pi}; \vec{q})
&=
\langle \mathrm{O}_{B^*_\mu}(\vec{q}, t) \mathcal{V}_{\mu}(\vec{q}, \tau)  \rangle_F
\\
&\times
\langle \mathrm{O}_{\pi}(-\vec{q}, t) \mathrm{O}^\dagger_\pi(-\vec{q}, 0)  \rangle_F
\end{align}
which predicts that the current-enhanced $B^*_\mu \pi$ states are in the momentum configuration where the pions in
the final and initial state carry the same momentum, and $B^*$ carries the same momentum as the current.
In fact, in \cite{Bar:2023sef}, it was predicted in Heavy Meson ChPT, that such states contribute largely to this channel.

Similar discussions hold for other processes like $B_s \to K \ell \nu_\ell$, investigated, for instance, in \cite{FermilabLattice:2019ikx}. 
In this case, the largest excited-state contamination is expected to come from $B^* K$ states.
These examples illustrate how the mechanism discussed here applies beyond nucleon observables, and the framework can be naturally extended to a broad class of hadronic transitions.

{\textbf{Conclusions.---}}%
Excited-state contamination remains a leading source of systematic uncertainty in lattice QCD determinations of hadron structure.
We have identified a key mechanism behind this effect: external currents can couple strongly to specific multi-particle states whose contributions are not volume-suppressed, leading to an effective current-enhancement of these states.
Such current-enhanced states—often overlooked in standard analyses for lack of direct ChPT predictions—can dominate signals at the source–sink separations accessible today.
We present a general framework, supported by numerical evidence, that determines which excited states contribute most significantly and why.
This picture agrees with existing chiral effective theory results, but extends beyond it by providing a diagrammatic criterion based on the Wick contraction structure, applicable to any channel.
It clarifies, for example, why $N\pi$ states are not the dominant excited states in certain channels—such as the scalar channel—and thus why their inclusion in \cite{Alexandrou:2024tin} yielded improvements only where current-enhanced diagrams contribute, while having little effect elsewhere.
In practical terms, our findings motivate the targeted inclusion of current-induced multi-hadron states in multi-state fits, and provide a clear operator-selection criterion for variational analyses.
Adopting these strategies can systematically reduce excited-state systematics, paving the way for higher-precision determinations of hadron structure in future lattice QCD calculations.

{\textbf{Acknowledgments.---}}%
The author gratefully acknowledges illuminating discussions and collaboration with G. Bali and S. Collins, as well as valuable input from J. Green, R. Gupta, A. Patella, S. Prelovšek, S. Schaefer, and R. Sommer.
Special thanks are due to G. Bali, J. Green and K.-F. Liu for comments on an earlier version of this manuscript, and to C. Alexandrou and Y. Li for sharing their data and further confirming the volume scaling of quark-line disconnected diagrams.
This work was supported by the German Research Foundation (DFG) through the research unit FOR5269 “Future methods for studying confined gluons in QCD.”
Simulations were carried out on the QPACE 3 computer of SFB/TRR-55, using an adapted version of the \textsc{Chroma} \cite{Edwards:2004sx} software package.

\bibliography{main}
\bibliographystyle{unsrt}

\newpage
\appendix
\section{Topologies of Wick contractions}
\onecolumngrid

\begin{figure}[htbp]
  \centering
  \includegraphics[width=0.43\textwidth]{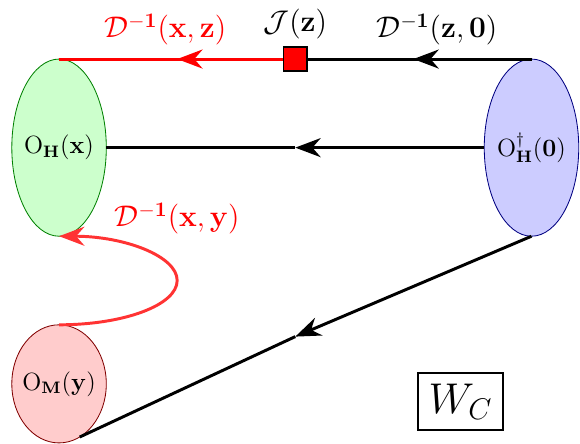}%
  \hfill
  \includegraphics[width=0.43\textwidth]{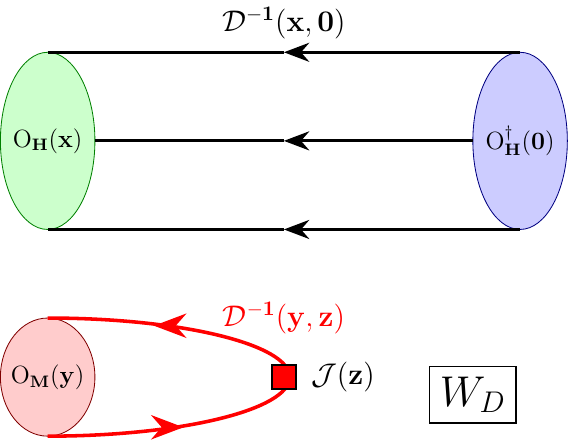}
  \caption{Examples of quark-line connected (left) and direct (right) diagrams arising from Wick contractions of baryon-to-baryon-meson three-point functions.}
  \label{fig:wick}
\end{figure}

\twocolumngrid
\section{Nucleon isovector vector matrix elements with a variational analysis}
\label{app:vector_varanalysis}
We study nucleon vector matrix elements using an effective variational basis consisting of nucleon and nucleon-vector-meson ($N\rho$-like) operators. 
The $N\rho$-like operators are constructed to have nucleon quantum numbers via the lattice group theory projection method discussed in \cite{Prelovsek:2016iyo}.
For zero total momentum, the explicit $N\rho$ operator with unit relative momentum is given in \cite{Prelovsek:2016iyo}; for unit total momentum we employ $\mathrm{O}^{(1)}_{N\rho}(\vec{1}) \sim \mathrm{O}_N(\vec{0})\mathrm{O}_\rho(\vec{1})$, and $\mathrm{O}^{(2)}_{N\rho}(\vec{1})\sim\mathrm{O}_N(\vec{1}) \mathrm{O}_\rho(\vec{0})$.
We form two operator bases: one with $\mathrm{O}_N$ and $\mathrm{O}^{(1)}_{N\rho}(\vec{1})$, and the other one with $\mathrm{O}_N$ and $\mathrm{O}^{(2)}_{N\rho}(\vec{1})$, to study respectively the forward and off-forward vector matrix elements.
Using this operator basis, we construct the matrix of two-point functions:
\begin{equation}
    C_{\rm 2pt}(\vec{p}, t)_{ij} = \langle \mathrm{O}_i(\vec{p}, t) ~\mathrm{O}_j^\dagger(\vec{p}, 0) \rangle~
\end{equation}
and matrix of three-point functions:
\begin{align}
C_{\rm 3pt}(\vec{p}', t; \vec{q}, \tau) = \langle \mathrm{O}_i(\vec{p}', t) ~\mathcal{J}(\vec{q}, \tau) ~ \mathrm{O}_j^\dagger(\vec{p}, 0)\rangle~,
\end{align}
with $\mathrm{O}_{i, j} \in \mathbb{B}_N=\left\{ \mathrm{O}_N, \mathrm{O}_{N \rho} \right\}$.
Like in \cite{Barca:2024hrl, Barca:2022uhi}, we neglect the computation of the three-point correlation function involving $\mathrm{O}_{k}=\mathrm{O}_{j}=\mathrm{O}_{N\rho}$,
because it is expected to be suppressed with the volume compared 
to the off-diagonal ones as no current-enhanced diagram contributes.

The quark fields within the $N$-like and $N\rho$-like operators are smeared both in the same way such 
that the nucleon smearing radii is $\approx 0.3~\rm fm$. This smearing procedure is carried out to enhance 
the overlap of the $N$- and $N\rho$-like operators with the states $N$ and $N\rho$, respectively.

We solve the Generalised EigenValue Problem (GEVP)
\begin{equation}
C_{\rm 2pt}(\vec{p}, t) V(\vec{p}, t_0) = C_{\rm 2pt}(\vec{p}, t_0) \Lambda(\vec{p}, t) V(\vec{p}, t)
\qquad t>t_0~,
\end{equation}
for $\vec{p}=\vec{0}, \vec{1}$.
When solving in the unit-lattice-momentum frame, the effective masses of the first and second eigenvalues lie close to the $N$ and non-interacting $N\rho$ energies (Fig.~\ref{fig:gevp_ratios}).
In contrast, for the zero-momentum frame we are unable to resolve the $N\rho$ state, as the signal decays to well below the expected $N\rho$ energy before being overwhelmed by noise.
Notably, there are several states below the $N\rho$, both at zero and total momentum, 
namely $N\pi$, $N\sigma$, and $N\pi\pi$. 
{
\begin{figure*}[t!]
  \centering
    \includegraphics[width=\textwidth]{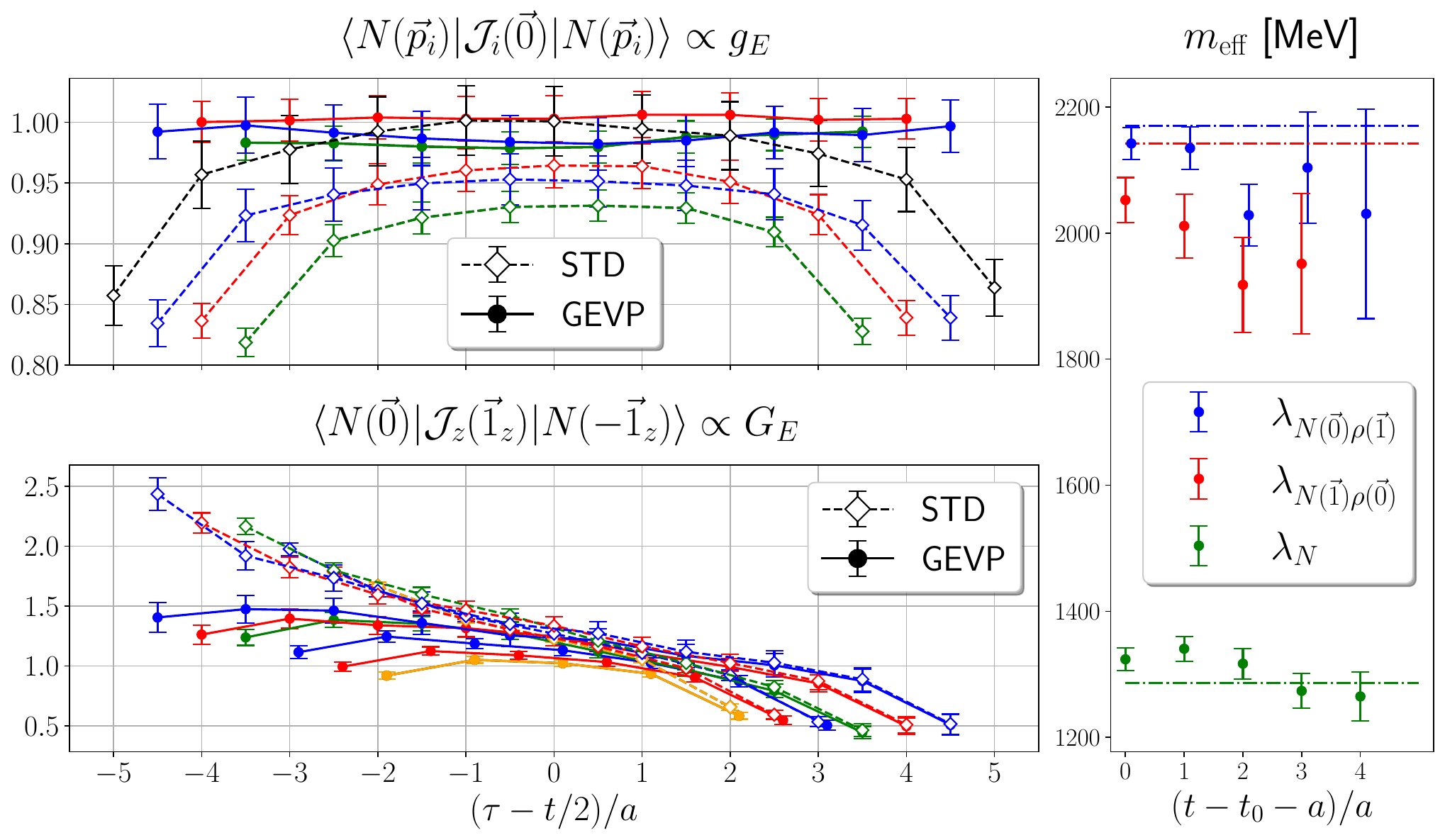}
    \caption{(left) Comparison between standard ratios and GEVP-improved ratios in the forward limit (top) and off-forward limit (bottom). In both cases, the effective variational analysis reduces exponentially the excited states.
    (right) Effective energies of the eigenvalues in the moving frames (data points) and non-interacting energies of $N$, $N(\vec{1})\rho(\vec{0})$, and $N(\vec{0})\rho(\vec{1})$. The energies used for estimating the non-interacting levels are computed on much higher statistics. 
    There are other states below or near the $N\rho$ states, namely $N\pi$, $N\pi\pi$, and $N\sigma$, see \cite{Barca:2024hrl, Barca:2022uhi}.
    The GEVP solutions are obtained for $t_0=3a$.}
  \label{fig:gevp_ratios}
\end{figure*}
}

Using the eigenvector components relative to the GEVP state $N$,
we subtract the $N\rho$ contribution by constructing the linear combination of GEVP-improved three-point functions
\begin{align}
\widetilde{C}&_{\rm 3pt}(\vec{p}', t; \vec{q}, \tau)
=\\ \nonumber
&\sum_{\mathrm{O}_{i, j} \in \mathbb{B}_N}
v_{\mathrm{O}_i}^{N}(\vec{p}')~
\langle \mathrm{O}_i(\vec{p}', t)~\mathcal{J}(\vec{q}, \tau)
\mathrm{O}^\dagger_j(\vec{p}, 0)\rangle
~v_{\mathrm{O}_j}^{N}(\vec{p})~.
\end{align}
Similarly, we construct GEVP-improved two-point functions like
\begin{equation}
\widetilde{C}_{\rm 2pt}(\vec{p}, t)
=
\sum_{\mathrm{O}_{i} \in \mathbb{B}_N}
v_{\mathrm{O}_i}^{N}(\vec{p})~
\langle \mathrm{O}_i(\vec{p}, t)
~\mathrm{O}^\dagger_i(\vec{p}, 0)\rangle
~v_{\mathrm{O}_i}^{N}(\vec{p})~.
\end{equation}

In the forward limit with $\vec{p}' = \vec{p} = \vec{1}$, the nucleon vector three-point functions are significantly affected by excited-state contamination, as also reported in \cite{Alexandrou:2024tin} for both physical and heavier-than-physical pion masses. This channel is particularly interesting because the corresponding matrix element is proportional to the nucleon electric charge. To suppress contributions from $N\rho$ intermediate states, we perform a variational analysis using a basis of operators $\mathrm{O}_N(\vec{1})$ and $\mathrm{O}_N(\vec{1}) \mathrm{O}\rho(\vec{0})$.

In the off-forward case, substantial excited-state effects are likewise observed in certain channels \cite{Alexandrou:2024tin}. We focus on the kinematic setup $\vec{p}' = \vec{0}$, $\vec{q} = -\vec{p} = \vec{1}_k$ (unit lattice momentum along direction $k$), and $\mathcal{J}_\mu = \mathcal{J}_k$.
Here we use the basis ${\mathrm{O}_N(\vec{1}), \mathrm{O}_{N \rho}^{(1)}(\vec{1})}$ to reduce $N\rho$ contamination.

The choice of these operator bases is motivated by the fact that they yield non-vanishing quark-line disconnected (direct) diagrams, which dominate the signal up to Euclidean time separations of approximately $\tau \approx 0.7~\mathrm{fm}$.

In Fig.~\ref{fig:gevp_ratios}, we present a comparison between standard and GEVP-improved ratios in both the forward (top) and off-forward (bottom) kinematics. The impact of including $N\rho$ operators is evident: the dependence on the source-sink separation $t$ and interaction time $\tau$ is significantly reduced. In the off-forward case (bottom plot), the remaining contamination at $\tau \gtrsim t/2$ can be attributed mainly to $N(-\vec{1})\rho(\vec{1})$ states, which are not included in the variational basis.

The analysis is performed on a single ensemble with $400$ gauge configurations, $m_\pi = 429~\mathrm{MeV}$, $L = 24a \approx 2.4~\mathrm{fm}$, and $N_f = 3$, which is a subset of the ensemble used in previous variational studies \cite{Barca:2024hrl, Barca:2022uhi}. This ensemble was selected because the $\rho$ meson has a mass of $m_\rho \approx 860~\mathrm{MeV}$, well below the lowest two-pion P-wave energy $E_{\pi(\vec{p})\pi(-\vec{p})}$. This allows us to clearly identify the meson and directly connect with the discussion in the main text.

At lighter pion masses, we expect the dominant excited-state contributions to arise from three-particle states such as $N(\vec{0})\pi(\vec{0})\pi(\vec{p})$ and $N(\vec{0})\pi(-\vec{p})\pi(\vec{p})$ in off-forward matrix elements, and from $N(\vec{0})\pi(\vec{p})\pi(\vec{p})$ in the forward limit relevant for nucleon vector charges. These states are not the lowest in the spectrum and can therefore be easily missed in conventional multi-state fits.

\end{document}